\documentclass{article}

\usepackage[preprint,nonatbib]{neurips_2019}

\usepackage[utf8]{inputenc} 
\usepackage[T1]{fontenc}    
\usepackage{url}            
\usepackage{booktabs}       
\usepackage{amsfonts}       
\usepackage{nicefrac}       
\usepackage{microtype}      
\usepackage{pgfplots}
\pgfplotsset{compat=1.14}
\usepackage{graphicx}
\usepackage{subfig}
\usepackage{amsmath}
\usepackage{booktabs} 
\newlength\figureheight 
\newlength\figurewidth 
\usepackage{tikz}
\usepackage{pgfplots}
\usepackage{mathtools}
\usetikzlibrary{positioning}

\DeclareMathOperator*{\minimize}{minimize}

\title{Perceptual Based Adversarial Audio Attacks}

\author{%
  Joseph Szurley$^1$ \\
  $^1$ \text{Bosch Center for Artificial Intelligence}\\
  Pittsburgh, PA 15222 \\
  \texttt{joseph.szurley@bosch.com} \\
  \And
    J. Zico Kolter$^{1,2}$\\
$^2$ \text{Computer Science Department} \\ 
Carnegie-Mellon University \\
Pittsburgh, PA 15213\\
\texttt{zkolter@cs.cmu.edu}
}

\begin{document}

\maketitle

\begin{abstract}
Recent work has shown the possibility of adversarial attacks on automatic speech recognition (ASR) systems. However, in the vast majority of work in this area, the attacks have been executed only in the digital space, or have involved short phrases and static room settings. In this paper, we demonstrate a physically realizable audio adversarial attack. We base our approach specifically on a psychoacoustic-property-based loss function, and automated generation of room impulse responses, to create adversarial attacks that are robust when played over a speaker in multiple environments. We show that such attacks are possible even while being virtually imperceptible to listeners.
\end{abstract}
\section{Introduction}
While machine learning algorithms have shown impressive performance on a variety of tasks in recent years, it has been shown that these algorithms can succumb to so-called \textit{adversarial attacks} \cite{42503,Kurakin2017AdversarialEI,Goodfellow2015ExplainingAH}.  These adversarial attacks have been applied to a variety of tasks including image \cite{42503}, speech \cite{8424625,Schoenherr2019}, and control \cite{46154,10.1007/978-3-319-62416-7_19} domains.  In adversarial attacks, a small perturbation (often imperceptible to a human observer) is added to the input signal with the goal of changing the output of the algorithm.  This not only raises concerns over edge cases that may lead to unexplainable failures but also the security of systems that employ these algorithms.   

With the ubiquity of personal voice assistants (Siri, Alexa, Google Home, etc.) which rely on automatic speech recognition (ASR),  it is a natural question to ask how robust these systems are to adversarial attacks.  Typical adversaries for ASR systems are generated through \textit{targeted attacks} which change the characters in the original utterance such that a target phase is transcribed.  While there are instances where a single character or word replacement may be sufficient, for the majority of adversarial attacks on ASR systems the whole phrase needs to be attacked.  Furthermore, the adversarial attacks must be robust to perturbations so that they can be played over-the-air.    
 
In the audio domain, the psychoacoustic properties of the human auditory system can be exploited to generate attacks that are more potent yet less perceptible.  These rely on masking effects, where certain frequencies are less perceptible when presented simultaneously to the auditory system \cite{de023eec970e4fc28e5413ca9ed20f8f}.  Another benefit of relying on perceptual based measures is that any non-speech audio in the environment can be utilized to inject noise into the signal, i.e., more noise allows for stronger attacks.  This was utilized in \cite{217607} where the attack was based on a music piece that contained an adversarial speech command.   

In \cite{Yakura2018RobustAA} over-the-air attacks were successfully implemented by incorporating typical transformations in the adversarial generation process that an audio signal encounters from source to receiver.  However, the adversarial generation did not incorporate any psychoacoustic properties which produced audible distortions to the signal.  These psychoacoustic properties were included in \cite{2019arXiv190310346Q} but relied on back-propagation through the frequency domain and were not carried out in an over-the-air scenario.  Furthermore, most previous adversarial audio work relies on subjective human listening tests and metrics that do not explicitly account for the perceptual distortion of the adversarial signal, e.g., only the SNR difference is used.

In this paper, we propose a novel formulation of a psychoacoustic based attack for adversarial audio generation.  This formulation exploits the psychoacoustic masking thresholds of the original signal and has an explicit trade-off between the audio distortion and strength of the adversary.  The adversary can be generated completely in the time domain (once the psychoacoustic properties of the original phrase are converted from the frequency domain).  We evaluate the adversarial audio using the Perceptual Evaluation of Speech Quality (PESQ) score, which eliminates the need for human listening tests.  Finally, we perform over-the-air tests to highlight the robustness of the adversaries.  
\section{Background and Related Work}
Typical deep learning based ASR systems are end-to-end models which take raw speech segments and output a transcript.  Since the processing of the audio signal is a multi-step process, attacks can target either the raw signal or the signal after it has been processed in a transform domain, e.g. after a Fast Fourier Transform (FFT) operation.  To further increase the potency of the attacks  they can be applied in such a way to exploit properties of the signal in both the time and frequency domains.    

In \cite{carlini2016hidden,vaidya2015cocaine,DBLP:journals/corr/abs-1801-00554}, attacks were generated by manipulating the Mel-frequency cepstral coefficients (MFCC) and \textit{inverse transforming} the coefficients back to the time domain.  However, due to the lossy nature of the MFCC computation, some of the information of the original audio signal is lost, limiting the attack space.  The raw signal was attacked in \cite{8424625,Yakura2018RobustAA} in a complete end-to-end fashion.  These attacks focused on limiting the size of the adversary based on the signal-to-noise ratio (SNR) only, and did not account for the perceptual distortion introduced by the adversary.    

In \cite{Yakura2018RobustAA} these attacks were extended to incorporate typical transformations in the adversarial generation process that an audio signal encounters from source to receiver.  This was shown to improve the robustness of the attacks for over-the-air broadcasting but were more perceptible to listeners.  Psychoacoustic properties were used to reduce the perceptibility of the adversarial signal in \cite{Schoenherr2019,2019arXiv190310346Q}, but both methods encountered instability during back-propagation in the frequency domain.  These methods were also not explored in the physical domain, i.e., they were conducted in the pure digital domain.  Due to path effects from the speaker to the receiver, as well as additive noise, the success of the adversary may be severely affected in an over-the air scenario.    

We propose a formulation for psychoacoustic based attacks that does not have the aforementioned stability issues in the frequency domain.  The formulation exploits the symmetry properties of the Discrete Fourier Transform (DFT) so that, once the relevant perceptual measures are extracted from the original audio, the attack can be performed solely in the time domain.  This reduces the memory requirements of the attack, compared to a frequency domain formulation, as there is no need to store both real and imaginary components.  We further add robustness to the attacks for over-the-air conditions by using generated room impulse responses.   

For perceptual evaluation of the adversarial signals, we use the Perceptual Evaluation of Speech Quality (PESQ) score \cite{Rix:2001:PES:1258236.1259107} which has a range from 0.5 (poor) to 4.5 (excellent).  The PESQ score is an objective measure to assess the speech quality of a system that induces a distortion on a reference speech signal in a completely automatic fashion, i.e., no listeners are necessary.  It therefore helps characterize the quality of the speech signal as it would be perceived by a listener.  

The generated attacks are shown to have high perceptual scores, i.e. low audible distortion, while still changing the transcription of the ASR system.  The adversarial attacks are also shown to be robust when played over a speaker in mono (original audio and adversary on the same channel) and stereo (original audio and adversary on different channels) fashion.   
\section{Methodology}
\subsection{Model Architecture}
We use the DeepSpeech \cite{Hannun2014DeepSS} model to generate adversarial audio examples.  DeepSpeech is a speech-to-text multi-layer bi-directional model that uses the Connectionist Temporal Classification (CTC) loss \cite{Graves}.  A raw audio waveform $x$ is fed to the DeepSpeech model to produce a character level probability for each frame, which when decoded, produces an output phrase $y$.  The CTC loss function is further able to score the most probable characters and phrases through different time alignments of $x$.  The CTC is a fully differentiable function which can therefore be exploited to allow for adversarial generation on a per character level through each time instance over the entire length of $x$.  
\subsection{Adversarial Audio Generation}
In adversarial audio attacks, a perturbation, $\delta$, is applied to the original raw waveform $\tilde{x} = x +\delta$ such that the output is now changed to target phrase $y_t$.  This can be formulated as an optimization problem over $\delta$ of the form \cite{8424625}
\begin{align}
\label{equation:original_loss}
\minimize_{\|\delta\|_2 \le \epsilon} \; L(x+\delta,y_{target})
\end{align}
where $L(\cdot)$ is the loss function and where $\|\cdot\|_2$ is an $l_2$-norm.  This minimization problem is solved over the complete audio signal, again by exploiting the CTC loss function, ensuring that the $l_2$-norm of the adversary is inside some $\epsilon$-ball.

The adversary in (\ref{equation:original_loss}) is only constrained to be inside an $\epsilon$-ball and is usually chosen in such a way as to minimally perturb the input.  However, even if the attack is successful using this formulation, audible artifacts may still be perceptible.  This is a result of the addition of certain frequency components, when considering the attack from the frequency domain, where the human auditory system has a varying sensitivity to intensity as a function of frequency.  This sensitivity can be included in the optimization in (\ref{equation:original_loss}) and furthermore exploited, to better \textit{mask} the adversary.  
\subsection{Psychoacoustic Model}
\label{subsection:Psychoacoustic_Model}
The sensitivity of human auditory system is a function of both intensity, typically measured with a logarithmic sound pressure level (dB SPL), and frequency.  It does not have a uniform response, requiring  as little as -5 dB SPL (light leaf rustling) in the peak regime (2-5 kHz), and requires higher intensities, especially as the bounds of human hearing are approached ([20Hz, 20kHz]).

Due to this sensitivity and the discrete way in which sound is processed by the auditory system, a masking effect occurs around \textit{critical bands} when multiple frequencies are presented simultaneously to a listener.  The critical bands can be thought of as drowning out other frequencies in the neighborhood, which is again both a function of frequency and intensity, i.e., low intensities produce smaller masking effects \cite{de023eec970e4fc28e5413ca9ed20f8f}.  This masking can therefore be exploited to embed the adversarial signal under a certain hearing threshold thereby ensuring that it remains imperceptible.  

Speech can be thought of as a dynamically changing process throughout the temporal domain.  In order to get an accurate representation of the frequency components, analysis is normally carried out on short segments, or frames.  Frame lengths are typically on the order of 10-20ms for speech processing [p. 41]\cite{Tashev:2009:SCP:1572522}, where it is assumed that the frequency components are stationary within this time frame.  There is a small amount of overlap between frames, to ensure frequency continuity, and a window functioning is applied to smooth the transition of frequency components.  Using this approach, the raw waveform $x$ is segmented into $N$ frames of length $L$ given as
\begin{equation}
\label{eq:window}
x_n (kT) = x(kT+nL)w_L(t-nL) \; k \in [0,N-1]
\end{equation}
where $n$ is the frame index and $w_L$ is a window function. 

The psychoacoustic model used to find the \textit{global masking threshold} is based on the MPEG-ISO \cite{ISO11172} and was included in the attack presented in \cite{Schoenherr2019}.  We will not explain in detail how the global masking threshold is generated and we refer the reader to \cite{842996,6255767,8424625} for a more in depth explanation. Calculating the global masking threshold per frame consists of the following 5 steps:
\begin{enumerate}
    \item The frame is first normalized to a standard dB SPL which converts each frame to roughly the same intensity levels.  While this is only an approximation of the dB SPL it is needed as signals have intensity levels that are functions of room dynamics, microphone responses and so forth.  The signal is then windowed and transformed to the FFT domain to generate a power spectral density (PSD).    
    \item Tonal and non-tonal maskers are then identified in the PSD.  The tonal maskers represent exact frequency peaks in the PSD while the non-tonal maskers are found by a geometric mean across a group of frequencies.  These maskers then help identify which frequencies become less perceptible when presented simultaneously.  
    \item The number of maskers is then reduced, or decimated, by comparing the tonal and non-tonal maskers using a sliding window scheme.  This reduces the granularity of the maskers and results in a smoother transition between peaks in the PSD.  
    \item The tonal and non-tonal maskers are then used to generate a \textit{masking pattern} that encompasses the adjacent frequencies.  
    \item The global masking threshold is the then found by combining the masking patterns from the previous step.  This global masking threshold then represents a perceptual weighting that is based on the intensity and frequency components of the signal as well as the psychoacoustic properties of the human auditory system.  
\end{enumerate}
The resulting global masking threshold $t$ can then be found for each frame $N$ across all frequencies $f$, $t_n(f) \in[0, \frac{f_s}{2}]$, where $f_s$ is the sampling frequency.  

\begin{figure}[t]
\vskip 0.2in
\begin{center}
    \setlength\figureheight{0.3\columnwidth}
  	\setlength\figurewidth{0.8\columnwidth}
\input{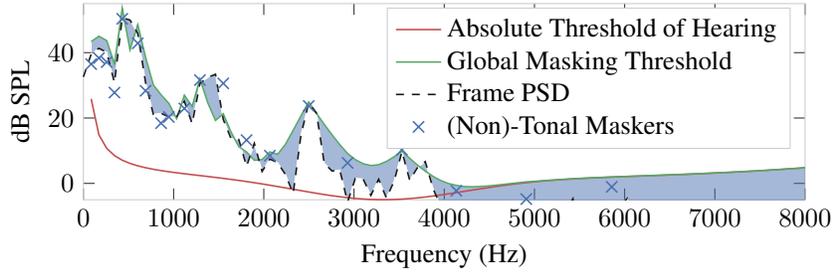}
\vspace{-3mm}
\caption{The absolute threshold of hearing compared to the global masking threshold extracted from a speech frame.}
\label{figure:spl}
\end{center}
\end{figure}

Figure \ref{figure:spl} shows the absolute threshold of hearing (the minimal dB SPL for a sound to be perceived), the tonal and non-tonal maskers and the global masking threshold for a single analysis frame.  The fill between the lines represents parts of the frequency band that will not be perceptible unless a higher dB SPL is reached, e.g., around 3 kHz there is a 10 dB SPL gap between the global masking threshold and the PSD of the analysis frame.  The adversarial signal can therefore be added to this portion of the frequency spectrum, up to the masking threshold, while remaining inaudible.
\subsection{Perceptual Based Audio Attack Optimization}
\label{ss:PBAAO}
Since the psychoacoustic model exploits the frequency domain relationship between the signal and the human auditory system, it is natural to first analyze how the attack can be formalized in the frequency domain.  Relying on the linearity property of the Fourier transform, the adversary at frame $n$ and frequency $f$ can be represented as
\begin{equation}
    \widetilde{X}_n(f) = {X}_n(f) + {\delta}_n(f)\;.
\end{equation}

The perceptual loss can be formulated as a weighting in the frequency domain [p.60]\cite{Oppenheim:1999:DSP:294797} between the original waveform and the adversarial waveform represented as
\begin{equation}
\label{eq:freq}
L_{percep}(\delta) = \frac{1}{2N} \sum_{n=0}^{N-1} \sum_{f=0}^{\frac{fs}{2}} w_{n}(f)|\delta_n(f)|^2
\end{equation}
where $w_n(f)$ is a perceptual based weighting factor.  This weighting factor is based on the global masking threshold derived in Section \ref{subsection:Psychoacoustic_Model} as
\begin{equation}
\label{equation:weighting}
w_n(f) = 10^{-\beta t_n(f)}
\end{equation}
where $\beta$ is a compression factor that defines how much of the perceptual based weighting should be included and has been heuristically determined to lie in the range [0.04-0.06] \cite{6255767}.  This can be thought of similarly to a power weighting between the original signal and the adversary.

The frequency based perceptual loss in (\ref{eq:freq}) can now be reformulated in the time domain as:
\begin{equation}
\label{equation:perceptual_loss}
L_{percep}(\delta) = \frac{1}{2N}\sum_{n=0}^{N-1} \delta_n^T D^H W_n D \delta_n  \equiv \frac{1}{2N}\sum_{n=0}^{N-1} \delta_n^T G_n \delta_n 
\end{equation}
where $(\cdot)^H$ represent the conjugate transpose of a matrix, $D$ is the DFT matrix, and $W_n$ is a symmetric weighting matrix around $\frac{fs}{2}$ defined as: 
\begin{equation}
\label{equation:cc}
W_n = \begin{bmatrix} 
w_n(0) &  0 & \ldots & 0 \\
0& w_n(1) & \ldots & 0  \\
\vdots & \vdots & \vdots & \vdots\\
0& \ldots & w_n(1)  & 0  \\
0& \ldots & 0 & w_n(0)  \\ 
\end{bmatrix}   \;.
\end{equation}
\newline
For the ease of exposition we represent the multiplication of the weighting matrix and the DFT matrices as $G_n = D^HW_nD$.  Note that because of the symmetry properties of the DFT, the imaginary components are cancelled during the multiplication such that $G_n \in \mathbb{R}^{L\times L}$,  where L is the DFT window length which is assumed to be equal to the frame length in (\ref{eq:window}).

We can now combine the perceptual based loss in (\ref{equation:perceptual_loss}) and the adversarial loss in (\ref{equation:original_loss}) as 
\begin{equation}
\label{eq:full}
\minimize_{\|\delta\|_2\leq \epsilon} \; \bigl( \alpha L(x+\delta, y_{target}) + (1-\alpha) L_{percep}(\delta) \bigr) \equiv \minimize_{\|\delta\|_2\leq \epsilon} \; L_{tot}(x+\delta, y_{target}, \alpha)
\end{equation}
where $0\leq \alpha \leq 1$ and $L_{tot}$ represents the total loss.  While $\epsilon$ constrains the overall size of the perturbation, $\alpha$ controls the amount of perceptual based penalty included in the optimization.

One of the benefits of using equation (\ref{eq:full}) is that the $G$ matrices need only be computed once before optimizing the adversary as the original audio is not changed, i.e., the global masking threshold remains constant throughout the optimization procedure.  This means that the adversarial signal does not need to be transformed to the frequency domain at each iteration.  Compared to previous attacks that take place in the frequency domain, this approach reduces the memory needed, as real and complex components are no longer stored, and alleviates the previously mentioned stability problems in the optimization.

We use the $l_2$ norm constraint on $\delta$ in (\ref{eq:full}) which allows for the attacks to be more \textit{localized} in the audio signals.  This is contrast to other adversarial audio attacks which use an $l_\infty$ allowing for equally large perturbations across the whole audio signal.  Since the adversary is now constrained based on the $l_2$-norm, it typically has larger values than that of an $l_\infty$ attack.  To solve (\ref{eq:full}) we rely of the projected gradient descent (PGD) method 
\begin{equation}
\label{eq:PGDstep}
    \delta \coloneqq \mathcal{P}_\epsilon\left(\delta - \mu  \frac{\nabla_\delta L_{tot}(x +\delta,y_{t},\alpha)}{\| \nabla_\delta L_{tot}(x+\delta,y_{t},\alpha)\|}\right)
\end{equation}
where $\mathcal{P}_\epsilon$ is the projection on the $\epsilon$ ball and $\mu$ is the step size.  The projection of (\ref{eq:PGDstep}) is then given as
\begin{equation}
    P_z = \epsilon\frac{z}{max\{\epsilon,\|z\|_2\}}\;.
\end{equation}
Before the projection, we also include the constraint $-1\leq x+\delta \leq 1$ to ensure that the adversarial signal lies within the valid normalized audio signal range.  The normalization of the gradient in (\ref{eq:PGDstep}) also helps stabilize the descent direction.  
\section{Experimental results}
The DeepSpeech model was trained in pyTorch using the Librispeech dataset \cite{7178964} which consists of 960 hours of clean audio with corresponding transcriptions.  During training, a sampling frequency of $f_s=16$ kHz was used and the data was augmented by adding random noise as well as pitch and time shifting the signal.  The compression factor in (\ref{equation:weighting}) was $\beta=0.06$ for all experiments.  The probabilities from the DeepSpeech model were decoded using a greedy method, i.e. at each instance, only the character with the highest probability is selected.  

In order to assess the performance of the attack, we used several metrics that analyzed both the signal characteristics and final transcription.  
The word error rate (WER) is derived from the Levenshtein Distance algorithm and defines the minimum edit between two strings given as:
\begin{equation}
\text{WER} = \frac{S+D+I}{N}
\end{equation}
where $S$ is the number of substitutions, $D$ is the number of deletions, $I$ is the number of insertions, and $N = S + D +C$ where $C$ is the number of correct words.  For a perfect attack $y=y_t$, $S=D=I=0$ producing a WER=0.  As the distance between the two string increases $y \neq y_t$, i.e. more characters and words are changed, the WER likewise increases.  The character error rate (CER) is the per-character difference between two strings and CER=0 when $y=y_t$.  For perceptual evaluation we used the Perceptual Evaluation of Speech Quality (PESQ) score \cite{Rix:2001:PES:1258236.1259107} which has a range from 0.5 (poor) to 4.5 (excellent).  The PESQ score was calculated in \textit{full reference} mode, which is a sample-by-sample distortion comparison between $x$ and $\tilde{x}$ after a temporal alignment.  The output SNR was estimated using the original signal and $\delta$ as the noise signal.  

Generated adversaries using the phrase "\textit{open the door}" as $y_t$ with an $\epsilon=1000$ had a 100\% success rate (WER = CER = 0) on 100 randomly sampled audio files from the Librispeech test set when no perceptual weighting was used ($\alpha=1$).  The same randomly sampled files were attacked again, this time with $\alpha=0.8$ allowing for perceptual weighing to be included in the loss function.  The perceptually weighted files again had a 100\% success rate (WER = CER = 0).

The SNR and PESQ scores were calculated for each signal using the perceptual ($\alpha=0.8$) and non-perceptual based ($\alpha=1$) attacks. In Table \ref{Table:compare}, we see that there is drop in the SNR and rise in PESQ score when using a perceptual based attack compared to that of a non-perceptual based attack.  This reason for this can be intuitively thought of as follows:  When no perceptual weighting is used, the attack can equally spread anywhere on the frequency spectrum.  This spreading in the frequency domain reduces the overall amplitude for any one frequency component which corresponds to a lower adversarial signal power and hence higher SNR.  However, when the perceptual loss is included, the attack is focused more around frequencies that dominate the global masking threshold.  This has the effect of increasing the power on some of the components, lowering the SNR, but ensuring that these lie within the masking threshold and are inaudible, raising the PESQ score.         
\begin{table}[t]
  \caption{Comparison of Perceptual and non-Perceptual Based Attacks on the LibiriSpeech Test Set} 
   \label{Table:compare}
  \centering
  \begin{tabular}{cccccc}
    \toprule
    Perceptual Weighting ($\alpha$)     & SNR (dB)     & PESQ & WER & CER\\
    \midrule
    0 & 27.9  & 3.3 & 0 & 0     \\
     0.8   & 24.1 & 4.0 &0 & 0     \\
    \bottomrule
  \end{tabular}
\end{table}
\begin{figure}[t]

\begin{center}
    \setlength\figureheight{0.25\columnwidth}
  	\setlength\figurewidth{0.7\columnwidth}
\begin{tikzpicture}

\definecolor{color0}{rgb}{0.298039215686275,0.447058823529412,0.690196078431373}

\begin{axis}[
        width=\figurewidth,
		height=\figureheight,
axis line style={black},
tick align=inside,
x grid style={black},
xlabel={$\alpha$},
xmajorticks=true,
xmin=0.3, 
xmax=1.0375,
xtick style={color=white!15.0!black},,
y label style={yshift=-2mm},
ylabel={PESQ},
ymajorticks=true,
ymin=2.5, 
ymax=4.5,
ytick style={color=black},
 x dir=reverse
]
\addplot [semithick, color0]
table {%
0.25 4.32457113265991
0.26 4.31489372253418
0.27 4.3170223236084
0.28 4.32099485397339
0.29 4.30532884597778
0.3 4.30601072311401
0.31 4.22004747390747
0.32 4.2104344367981
0.33 4.19467878341675
0.34 4.33440351486206
0.35 4.18880271911621
0.36 4.34831190109253
0.37 4.3442006111145
0.38 4.34142923355103
0.39 4.31932926177979
0.4 4.32283496856689
0.41 4.29357481002808
0.42 4.30665826797485
0.43 4.32047748565674
0.44 4.31198787689209
0.45 4.25217962265015
0.46 4.26925277709961
0.47 4.25436782836914
0.48 4.22288036346436
0.49 4.23124265670776
0.5 4.20404005050659
0.51 4.23583078384399
0.52 4.22411966323853
0.53 4.16796922683716
0.54 4.11603975296021
0.55 4.15473556518555
0.56 4.14019298553467
0.57 4.09210538864136
0.58 4.08890247344971
0.59 4.05265665054321
0.6 4.07751417160034
0.61 4.00093507766724
0.62 3.92007660865784
0.63 3.95999932289124
0.64 3.94774985313416
0.65 3.89024329185486
0.66 3.94623136520386
0.67 3.9473123550415
0.68 3.97000336647034
0.69 4.1136589050293
0.7 4.01092576980591
0.71 4.0006947517395
0.72 3.9713146686554
0.73 4.11288738250732
0.74 4.18255186080933
0.75 4.10627937316895
0.76 4.13523530960083
0.77 3.8469295501709
0.78 3.98275184631348
0.79 4.10586643218994
0.8 3.98755669593811
0.81 3.88318610191345
0.820000000000001 3.85782551765442
0.830000000000001 4.12771415710449
0.840000000000001 4.09474658966064
0.850000000000001 3.92133045196533
0.860000000000001 4.06659173965454
0.870000000000001 4.02031993865967
0.880000000000001 3.87121319770813
0.890000000000001 3.94658493995667
0.900000000000001 3.84904217720032
0.910000000000001 3.73570990562439
0.920000000000001 3.58901286125183
0.930000000000001 3.57084083557129
0.940000000000001 3.53069591522217
0.950000000000001 3.50663256645203
0.960000000000001 3.45602226257324
0.970000000000001 3.11960411071777
0.980000000000001 3.03414249420166
0.990000000000001 2.73288369178772
1 2.659836769104
};
\end{axis}

\end{tikzpicture}
\caption{PESQ score with varying $\alpha$ in (\ref{eq:full}).}
\label{figure:pesq}
\end{center}
\end{figure}
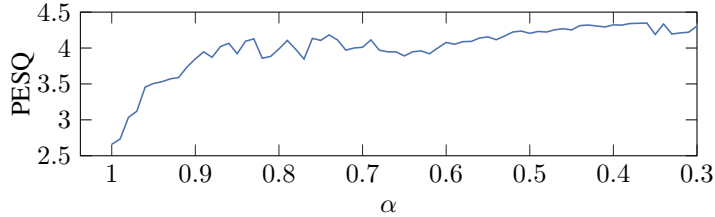

Figure \ref{figure:pesq} shows the PESQ score as $\alpha$ is decreased which was stopped when the CER $\neq 0$ at $\alpha=0.3$.  Surprisingly, the PESQ score rises rapidly with only a small decrease in $\alpha$, indicating a rapid rise in the perceptual quality.  With an $\alpha=0.4$, the PESQ score is almost at a maximal value of 4.5, so to a listener it would sound as if $x=\tilde{x}$.  It was observed that the attack was not always successful with very low values of $\alpha$.  This is most likely due to the adversary being heavily penalized for lying outside of the global masking threshold.  This limits the overall bandwidth the attack can use and may prevent it from changing enough of the signal to generate a successful attack.       

\begin{figure}[t]
\centering
  \setlength\figureheight{.1\columnwidth}
  \setlength\figurewidth{0.6\columnwidth}
\subfloat[][Original spectrogram]
{
	\begin{tikzpicture}
    \begin{axis} [%
        width=\figurewidth,
		height=\figureheight,
        scale only axis, 
        enlargelimits=false,
        ylabel style ={font=\footnotesize,yshift=-2mm},
        ylabel=Hz,
        clip=false,xticklabels=\empty,yticklabel style = {font=\small},yticklabels={0,0,,,8000}]
        \addplot graphics [
            xmin=0,
            xmax=15,
            ymin=0,
            ymax=6000,
        ] {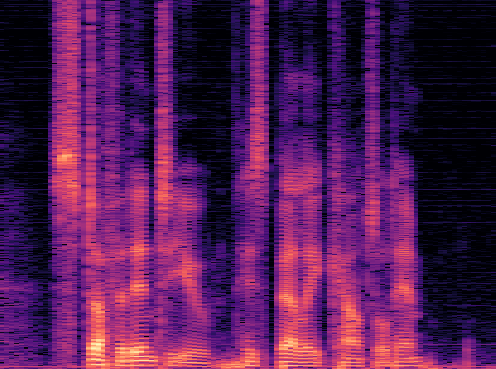};
    \end{axis}

\end{tikzpicture}
	\label{fig:org}
}\\ \vspace{-3mm}
\subfloat[Spectrogram with $\alpha=1$ in (\ref{equation:cc})]
{
	\begin{tikzpicture}
    \begin{axis} [%
        width=\figurewidth,
		height=\figureheight,
        scale only axis, 
        enlargelimits=false,
        ylabel style ={font=\footnotesize,yshift=-2mm},
        ylabel=Hz,
        clip=false,xticklabels=\empty,yticklabel style = {font=\small},yticklabels={0,0,,,8000}]
        \addplot graphics [
            xmin=0,
            xmax=15,
            ymin=0,
            ymax=6000,
        ] {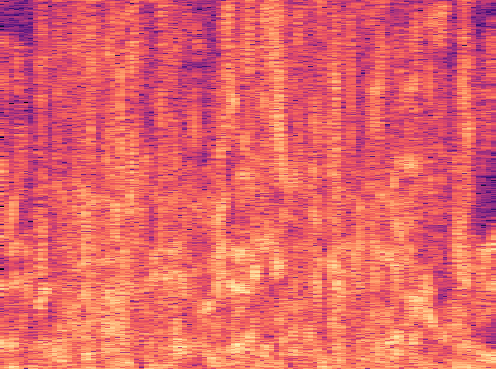};
    \end{axis}

\end{tikzpicture}
	\label{fig:alpha1}
}\\ \vspace{-3mm}
\subfloat[Spectrogram with $\alpha=0.5$ in (\ref{equation:cc})]
{
	\begin{tikzpicture}
    \begin{axis} [%
        width=\figurewidth,
		height=\figureheight,
        scale only axis, 
        enlargelimits=false,
        ylabel style ={font=\footnotesize,yshift=-2mm},
        ylabel=Hz,
        clip=false,xticklabels=\empty,yticklabel style = {font=\small},yticklabels={0,0,,,8000}]
        \addplot graphics [
            xmin=0,
            xmax=15,
            ymin=0,
            ymax=6000,
        ] {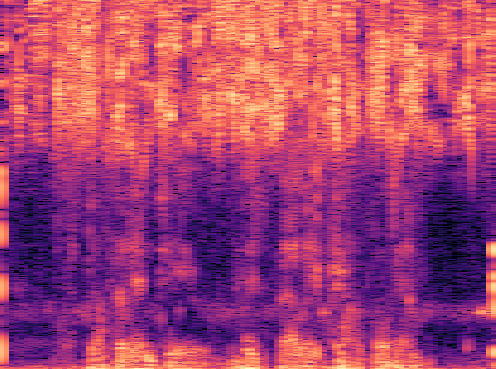};
    \end{axis}

\end{tikzpicture}
	\label{fig:alpha05}
	
}\vspace{-2mm}
\caption{Spectrogram of the original raw audio, $\alpha=1$ and $\alpha=0.5$.}
\label{fig:spec}
\vspace{-2mm}
\end{figure}

The effect of $\alpha$ can also be observed in the change of the spectrogram in Figure \ref{fig:spec}.  Figure \ref{fig:alpha1} shows the spectrogram when $\alpha=1$ so there is no constraint on where the adversary can attack.  We see that the adversary is spread almost evenly around all frequency bands.  In Figure \ref{fig:alpha05} , when $\alpha=0.5$, we see that the majority of the attack lies in these higher frequencies, especially when there is no speech present in the original signal which comes directly from the absolute threshold of hearing as shown in Figure \ref{figure:spl}.   
\subsection{Over-the-Air Attacks}
\label{subsec:oaa}
We further extended the testing to outside of the pure digital domain and perform over-the-air attacks (speaker/microphone combination).  In order to improve the robustness of signals for over-the-air attacks, we first pass them through a room simulator \cite{pyroomacoutsics} much like the Expectation over Transform method \cite{DBLP:conf/icml/AthalyeEIK18,2019arXiv190310346Q,Yakura2018RobustAA}.  Note that instead of using prerecording room impulse responses as in \cite{Yakura2018RobustAA}, we take a more robust approach as in \cite{2019arXiv190310346Q}, where a room simulator is used to generate responses with varying size and reflection coefficients for surfaces.  A bandpass filter was also applied in the range $f = [100,7500]$ to eliminate any spurious low frequency components and account for the frequency drop-off that occurs near the Nyquist rate in digital recording hardware.  

An Extech HD 6000 sound level meter was used to measure the loudness, using A-weighted decibels (dBA) which accounts for the sensitivity of human hearing, from the speaker to the microphone.  The over-the-air tests were performed in an anechoic chamber with a noise floor of 36 dBA.  The signals were broadcast using a Polk S20 speaker, Yamaha p2500s amplifier, and recorded using a MINI DSP UMIK-1 microphone. 

The microphone was first positioned close to the speaker to ensure a high SNR and eliminate any path effects between the microphone and speaker.  The distance between the microphone and speaker was then gradually increased, resulting in a lower SNR, to observe the effects of both added path effects and reduced signal power.  A language model decoder, based on the Librispeech 4-gram ARPA, was also added to the end of the DeepSpeech model in parallel to the greedy decoder.  The language model has the ability to error correct characters and words based on the speech corpus and, in some instances, can decrease the WER and CER rate.  

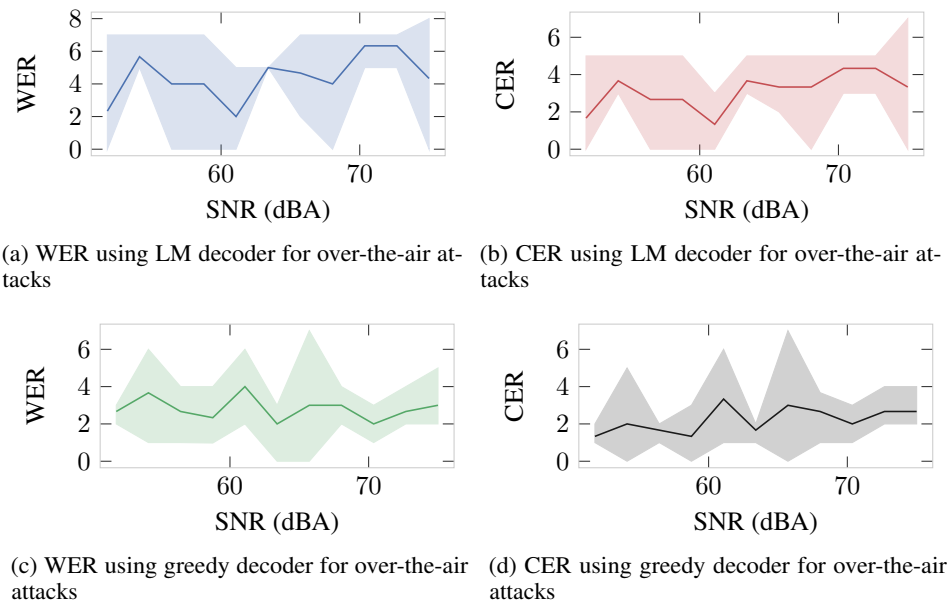
\begin{figure}[t]
\hspace{-5mm}
\centering
  \setlength\figureheight{.25\columnwidth}
  \setlength\figurewidth{0.45\columnwidth}
\subfloat[][WER using LM decoder for over-the-air attacks]
{
\begin{tikzpicture}

\definecolor{color0}{rgb}{0.298039215686275,0.447058823529412,0.690196078431373}

\begin{axis}[
width=\figurewidth,
		height=\figureheight,
axis line style={white!80.0!black},
tick align=inside,
x grid style={white!80.0!black},
xlabel={SNR (dBA)},
xmajorticks=true,
xmin=50.64, xmax=76.16,
xtick style={color=white!15.0!black},
y grid style={white!80.0!black},
ylabel={WER},
ymajorticks=true,
ymin=-0.4, ymax=8.4,
ylabel style ={yshift=2mm},
ytick style={color=white!15.0!black}
]
\path [draw=color0, fill=color0, opacity=0.2]
(axis cs:51.8,7)
--(axis cs:51.8,0)
--(axis cs:54.12,5)
--(axis cs:56.44,0)
--(axis cs:58.76,0)
--(axis cs:61.08,0)
--(axis cs:63.4,5)
--(axis cs:65.72,2)
--(axis cs:68.04,0)
--(axis cs:70.36,5)
--(axis cs:72.68,5)
--(axis cs:75,0)
--(axis cs:75,8)
--(axis cs:75,8)
--(axis cs:72.68,7)
--(axis cs:70.36,7)
--(axis cs:68.04,7)
--(axis cs:65.72,7)
--(axis cs:63.4,5)
--(axis cs:61.08,5)
--(axis cs:58.76,7)
--(axis cs:56.44,7)
--(axis cs:54.12,7)
--(axis cs:51.8,7)
--cycle;

\addplot [semithick, color0]
table {%
51.8 2.33333333333333
54.12 5.66666666666667
56.44 4
58.76 4
61.08 2
63.4 5
65.72 4.66666666666667
68.04 4
70.36 6.33333333333333
72.68 6.33333333333333
75 4.33333333333333
};
\end{axis}

\end{tikzpicture}
	\label{fig:w1}
} \hspace{2mm}
\subfloat[CER using LM decoder for over-the-air attacks]
{
\begin{tikzpicture}

\definecolor{color0}{rgb}{0.768627450980392,0.305882352941176,0.32156862745098}

\begin{axis}[
width=\figurewidth,
		height=\figureheight,
axis line style={white!80.0!black},
tick align=inside,
x grid style={white!80.0!black},
xlabel={SNR (dBA)},
xmajorticks=true,
xmin=50.64, xmax=76.16,
xtick style={color=white!15.0!black},
y grid style={white!80.0!black},
ylabel={CER},
ymajorticks=true,
ymin=-0.35, ymax=7.35,
ylabel style ={yshift=2mm},
ytick style={color=white!15.0!black}
]
\path [draw=color0, fill=color0, opacity=0.2]
(axis cs:51.8,5)
--(axis cs:51.8,0)
--(axis cs:54.12,3)
--(axis cs:56.44,0)
--(axis cs:58.76,0)
--(axis cs:61.08,0)
--(axis cs:63.4,3)
--(axis cs:65.72,2)
--(axis cs:68.04,0)
--(axis cs:70.36,3)
--(axis cs:72.68,3)
--(axis cs:75,0)
--(axis cs:75,7)
--(axis cs:75,7)
--(axis cs:72.68,5)
--(axis cs:70.36,5)
--(axis cs:68.04,5)
--(axis cs:65.72,5)
--(axis cs:63.4,5)
--(axis cs:61.08,3)
--(axis cs:58.76,5)
--(axis cs:56.44,5)
--(axis cs:54.12,5)
--(axis cs:51.8,5)
--cycle;

\addplot [semithick, color0]
table {%
51.8 1.66666666666667
54.12 3.66666666666667
56.44 2.66666666666667
58.76 2.66666666666667
61.08 1.33333333333333
63.4 3.66666666666667
65.72 3.33333333333333
68.04 3.33333333333333
70.36 4.33333333333333
72.68 4.33333333333333
75 3.33333333333333
};
\end{axis}

\end{tikzpicture}
	\label{fig:w2}
} \\ 
\subfloat[WER using greedy decoder for over-the-air attacks]
{
\begin{tikzpicture}

\definecolor{color0}{rgb}{0.333333333333333,0.658823529411765,0.407843137254902}

\begin{axis}[
width=\figurewidth,
		height=\figureheight,
axis line style={white!80.0!black},
tick align=inside,
x grid style={white!80.0!black},
xlabel={SNR (dBA)},
xmajorticks=true,
xmin=50.64, xmax=76.16,
xtick style={color=white!15.0!black},
y grid style={white!80.0!black},
ylabel={WER},
ymajorticks=true,
ymin=-0.35, ymax=7.35,
ylabel style ={yshift=2mm},
ytick style={color=white!15.0!black}
]
\path [draw=color0, fill=color0, opacity=0.2]
(axis cs:51.8,3)
--(axis cs:51.8,2)
--(axis cs:54.12,1)
--(axis cs:56.44,1)
--(axis cs:58.76,0.975000000000001)
--(axis cs:61.08,2)
--(axis cs:63.4,0)
--(axis cs:65.72,0)
--(axis cs:68.04,2)
--(axis cs:70.36,1)
--(axis cs:72.68,2)
--(axis cs:75,2)
--(axis cs:75,5)
--(axis cs:75,5)
--(axis cs:72.68,4)
--(axis cs:70.36,3)
--(axis cs:68.04,4)
--(axis cs:65.72,7)
--(axis cs:63.4,3)
--(axis cs:61.08,6)
--(axis cs:58.76,4)
--(axis cs:56.44,4)
--(axis cs:54.12,6)
--(axis cs:51.8,3)
--cycle;

\addplot [semithick, color0]
table {%
51.8 2.66666666666667
54.12 3.66666666666667
56.44 2.66666666666667
58.76 2.33333333333333
61.08 4
63.4 2
65.72 3
68.04 3
70.36 2
72.68 2.66666666666667
75 3
};
\end{axis}

\end{tikzpicture}
	\label{fig:w3}
}\hspace{2mm}
\subfloat[CER using greedy decoder for over-the-air attacks]
{
\begin{tikzpicture}

\begin{axis}[
width=\figurewidth,
		height=\figureheight,
axis line style={white!80.0!black},
tick align=inside,
x grid style={white!80.0!black},
xlabel={SNR (dBA)},
xmajorticks=true,
xmin=50.64, xmax=76.16,
xtick style={color=white!15.0!black},
y grid style={white!80.0!black},
ylabel={CER},
ymajorticks=true,
ymin=-0.35, ymax=7.35,
ylabel style ={yshift=2mm},
ytick style={color=white!15.0!black}
]
\path [draw=white!10.0!black, fill=white!10.0!black, opacity=0.2]
(axis cs:51.8,2)
--(axis cs:51.8,1)
--(axis cs:54.12,0)
--(axis cs:56.44,1)
--(axis cs:58.76,0)
--(axis cs:61.08,1)
--(axis cs:63.4,1)
--(axis cs:65.72,0)
--(axis cs:68.04,1)
--(axis cs:70.36,1)
--(axis cs:72.68,2)
--(axis cs:75,2)
--(axis cs:75,4)
--(axis cs:75,4)
--(axis cs:72.68,4)
--(axis cs:70.36,3)
--(axis cs:68.04,3.67499999999999)
--(axis cs:65.72,7)
--(axis cs:63.4,2)
--(axis cs:61.08,6)
--(axis cs:58.76,3)
--(axis cs:56.44,2)
--(axis cs:54.12,5)
--(axis cs:51.8,2)
--cycle;

\addplot [semithick, white!10.0!black]
table {%
51.8 1.33333333333333
54.12 2
56.44 1.66666666666667
58.76 1.33333333333333
61.08 3.33333333333333
63.4 1.66666666666667
65.72 3
68.04 2.66666666666667
70.36 2
72.68 2.66666666666667
75 2.66666666666667
};
\end{axis}

\end{tikzpicture}
	\label{fig:w4}
}
\caption{Word and character error rates for varying SNRs (3 tests at each level) using a language model (LM) and greedy (G) decoder.}
\label{fig:overtheair}
\vspace{-2mm}
\end{figure}

Figure \ref{fig:overtheair} shows the WER and CER for the over-the-air attack using both a greedy and language model decoder.   At a high SNR, the signal experiences clipping (the maximum amplitude of the signal is larger that the microphone response), which results in a high WER and CER (large string distance between $y$ and $y_t$).  Since the adversary is given as $x+\delta$, we see that when clipping occurs on the positive rail, $\max\{1,x+\delta\}$, this will result in a truncation of the adversary and, in the extreme case, truncation of the original signal.  The same will hold for the negative rail as well.  As the distance increases and the SNR lowers, moving the $x+\delta$ away from the rails,  the signal exhibits a lower WER and CER for both the language model (LM) and greedy decoder (G).  Between 60 and 70 dBA, the language model decoder had a WER=4.0 and CER=2.4 while the greedy decoder had a WER=3.0 and CER=2.25.

We further extended the over-the-air scenario to a two speaker setup which broadcasts the original audio and the adversary on two separate channels.  The speakers were separated by a distance of 4 inches and the microphone was placed 6 inches from the speakers which resulted in a $40^{o}$ separation between the speakers as observed from the microphone.  The signal was measured with an average of 66.7 dBA at the microphone during broadcasting.  Table \ref{Table:err_rate} shows the average error rates over 4 trials for a language model and greedy decoder.  The attack performs especially well using the language model decoder for the two speaker system, achieving a WER=0 and CER=0 for 3 of out the 4 trials.      
\begin{table}[t]
  \caption{Average error rates for two speaker over-the-air attack} 
   \label{Table:err_rate}
  \centering
  \begin{tabular}{cccc}
    \toprule
       & WER     & CER  \\
    \midrule
     Language Model Decoder & 0.5  & 0.5     \\
    Greedy Decoder & 5  & 4    \\
    \bottomrule
  \end{tabular}
\end{table}
\section{Conclusion}
This work has demonstrated a method for generating white-box adversarial examples with psychoacoustic based constraints.  The adversaries were generated in an end-to-end fashion which relied on extracting relevant psychoacoustic information from the original audio signal.  The optimization problem relied on an $l_2$-norm constraint in conjunction with the projected gradient descent method.  The perceptual quality was based on the PESQ score, which eliminated the need for exhaustive listening tests.  It was shown the adversarial examples typically lowered the SNR but, because of the psychoacoustic based constraints, the perceptual quality of the signal was increased resulting in a higher PESQ score.  Finally, the attacks were shown to be effective in over-the-air broadcasting.

In future work, transferability between architectures, similar to \cite{cisse2017houdini}, will be explored to understand the generalizability of the attacks.  While the current attacks are all performed in a white-box fashion (access to the entire model), black-box attacks will be explored to mimic more realistic scenarios.  Most ASR systems found in homes rely on microphone arrays and perform pre-processing steps such as, beamforming, noise reduction, etc.  The current attack does not account for these steps which will most likely severely impact the strength of the adversary.  The attacks also currently take several minutes to construct which is infeasible for real-time systems.  Future attacks will have to be computationally efficient while still retaining their robustness.      
\bibliographystyle{plain}
\bibliography{bibliography}
\end{document}